\documentclass[aps,floatfix,superscriptaddress,notitlepage,nofootinbib]{revtex4-1}
\usepackage{amsmath,amssymb,amsfonts,graphics,graphicx,dcolumn,bm,enumerate}
\usepackage{comment,natbib,appendix}
\usepackage{multirow,color}
\usepackage{chngpage}
\usepackage{afterpage}
\usepackage{xcolor}
\usepackage{amsthm}
\usepackage{natbib}
\usepackage{hyperref}
\usepackage[margin=1.2in]{geometry}
\usepackage{epstopdf}
\usepackage{float}
\usepackage{amsmath}
\usepackage{adjustbox}
\usepackage{booktabs}
\usepackage{cleveref}
\usepackage{hyperref}

\newcommand{\cmp}
{\affiliation{Saha Institute of Nuclear Physics, Kolkata 700064, India.}}
\newcommand{\isi}
{\affiliation{Economic Research Unit, Indian Statistical Institute, Kolkata 700108, India.}}
\newcommand{\raghunathpur}
{\affiliation{Department of Physics, Raghunathpur College, Raghunathpur, Purulia 723133, India.}}

\begin{document}
\title{Normalized fractional Hirsch Index and Modified Pareto Law}

\author{Asim Ghosh}
\email[Email: ]{asimghosh066@gmail.com}
\raghunathpur
 
 \author{Bikas K. Chakrabarti}%
 \email[Email: ]{bikask.chakrabarti@saha.ac.in}
 \isi  \cmp 
 
\begin{abstract}
The Hirsch index (introduced in 2005) has been a very successful scientometric author-level measure for the publication impact inequalities among the authors. However, it does not have any upper bound and statistically it grows in proportion to the square root of the number of publications by the author. Social scientists, on the other hand, measure the economic performance of different countries (having different wealth amounts and populations etc.) using inequality indices like the Gini index (since 1912) etc.,  which are normalized quantities  (independent of the country's total wealth or the population size). We show here that a normalized fractional Hirsch index, giving the fraction of total citations earned by fraction of those successful papers published by the author, can even indicate that typically about $14\%$ of the top cited papers earn about $86\%$ of citations for very successful scientists (like the 95 Nobel Laureates considered here). This is similar to Pareto's 80-20 law (since 1896) and agrees more precisely with those observed for self-organized sandpile models, where typically $86\%$  of the avalanche masses are dissipated through  $14\%$ of the largest avalanches as the piles approach their respective self-organized critical points.
\end{abstract}

\maketitle

\section{Introduction}
The Hirsch index ($h$) \cite{1Hirsch05}, introduced in 2005, is a scientometric author-level measure for both the productivity and citation impact of an author's publications. It is defined as the maximum value $h$ of the number of publications by the author such that each of them has been cited at least $h$ times in the literature. Though there are many criticisms (see e.g., \cite{2Koltun21}) the $h$-index is the most popular measure of scientific reputation of the authors, and both Google Scholar and Web of Science databases for the scientific authors provide it in their database. Hirsch index is a statistical (integer) measure of performance, based on the inequalities of the non-linear variations of citations among the publications by any scientist (higher citation papers are lower in number compared to those for the lower cited papers). The index has integer values and although it is generally higher for the successful authors, statistically it has been argued or observed to grow with the square root of the total number $C_t = \sum_p n_c$  of total citations of all the set of publications $p$ of the author (each having $n_c$ citations) \cite{11Yong14} or of total number of publications $N_t = \sum_p n_p$ of all the set of publications $p$ of the author (each of its $n_p$ papers having $n_c$  citations) \cite{12Ghosh23}. As such, the $h$-index does not have any normalizing upper bound value.

\begin{figure}[!tbh]
\centering
\includegraphics[width=1.\linewidth]{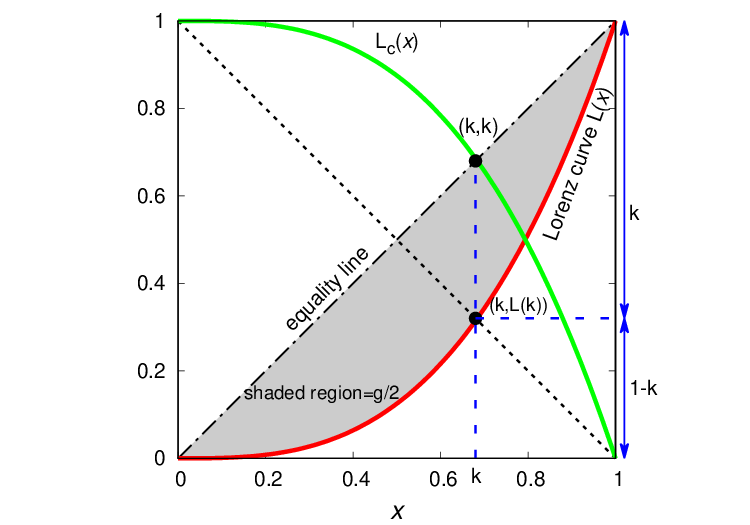}
\caption{ A schematic Lorenz curve or function ($L(x)$, in red) shows the proportion of total wealth (citation)  owned/possessed by a fraction ($x$) of people (papers) in ascending order of wealth (citations). The black dotted-dashed line represents a scenario of perfect equality. The Gini index $(g)$ is calculated from the area between the Lorenz curve and the equality line (shaded region), normalized by the total area (= 1/2) below the equality line.  The complementary Lorenz function
($L_c(x) \equiv 1 - L(x)$) is shown in green. The Kolkata index ($k$) is given by the fixed point of the complementary Lorenz curve: $L_c(k) = k$ or $L(k) = 1 - k$.}
\label{fig1}
\end{figure}

Socio-economic inequalities, specifically the income or wealth inequalities among different countries (say), on the other hand, are measured using normalized indices which are independent of their total wealth or the population. We will consider here specifically two such measures for comparison; namely the Gini index ($g$) \cite{3Gini1912} (since 1912), widely used still by social scientists (along with Pietra index since 1915 \cite{4Pietra15}) and the recently introduced (in 2014) one, called Kolkata index $k$ \cite{5Ghosh14} (see \cite{Ghosh26} for a recent study of the relationships between Gini, Pietra and Kolkata indices).   They are all normalized fractions and are bounded. Like the general nonlinear decrease in the number of higher-cited papers by any author, the number of people of any country decreases also nonlinearly with their wealth. Indeed, the Kolkata index value $k$ gives the fraction of wealth possessed by $(1 - k)$ fraction of the population ($k=0.8$ corresponds to Pareto's 80-20 law \cite{6Pareto1896} since 1896). A recent comparison \cite{7Banerjee23} of economic inequalities in the US (from IRS data for the period 1983-2018) indicates  $g = k \simeq 0.82$ and still has an increasing trend towards much higher inequality compared to the Pareto value $k = 0.80$. See also results there for citation inequalities ($ g = k \simeq 0.86$) for the publications by 20 Nobel laureates in physics, chemistry, medicine and economics.  Indeed it was shown already \cite{8Manna22}, the inequality  measures for the avalanche sizes (compared to their numbers) in the sandpile models like Bak-Tang-Weisenfeld and Manna models show $g = k \simeq 0.86$ just prior to the respective self-organized critical points.  Another recent study \cite{9Biswas25} for the citation inequality  (for 126,067 scientists, each having at least 100 publications, including 80  Nobel Laureates from the Google Scholar database) shows, while there is some general scatter for most scientists,  one gets $g = k \simeq 0.86$ with $g/k = 1.00 \pm 0.3$ for the Nobel Laureates, while their Hirsch index $h$ values scatter significantly (from 29 to 220).  This suggests the lack of any correlation of the Hirsch index $h$ with the scientific reputation of the respective scientists (compared to the correlations with the beyond-Pareto values of the Gini and Kolkata indices).

Here we propose that a simple normalization of the numbers associated with the Hirsch index $h$ can be found, called here the normalized fractional $h$-index $\bar{h}_c^{(f)}$, which will help us to discover major correlations with high-level inequalities. Similar to Pareto’s 80/20-like law, we will see that about 86\% citations come from 14\% of the top-cited papers by the ``top ranking" and ``successful" scientists. This is similar to that happening in sandpile models,  where \cite{8Manna22} typically  86\% of the avalanche masses are dissipated by 14\% of the largest avalanches near and above the self-organized critical point.

\begin{figure}[!tbh]
    \centering
    \includegraphics[width=0.7\linewidth]{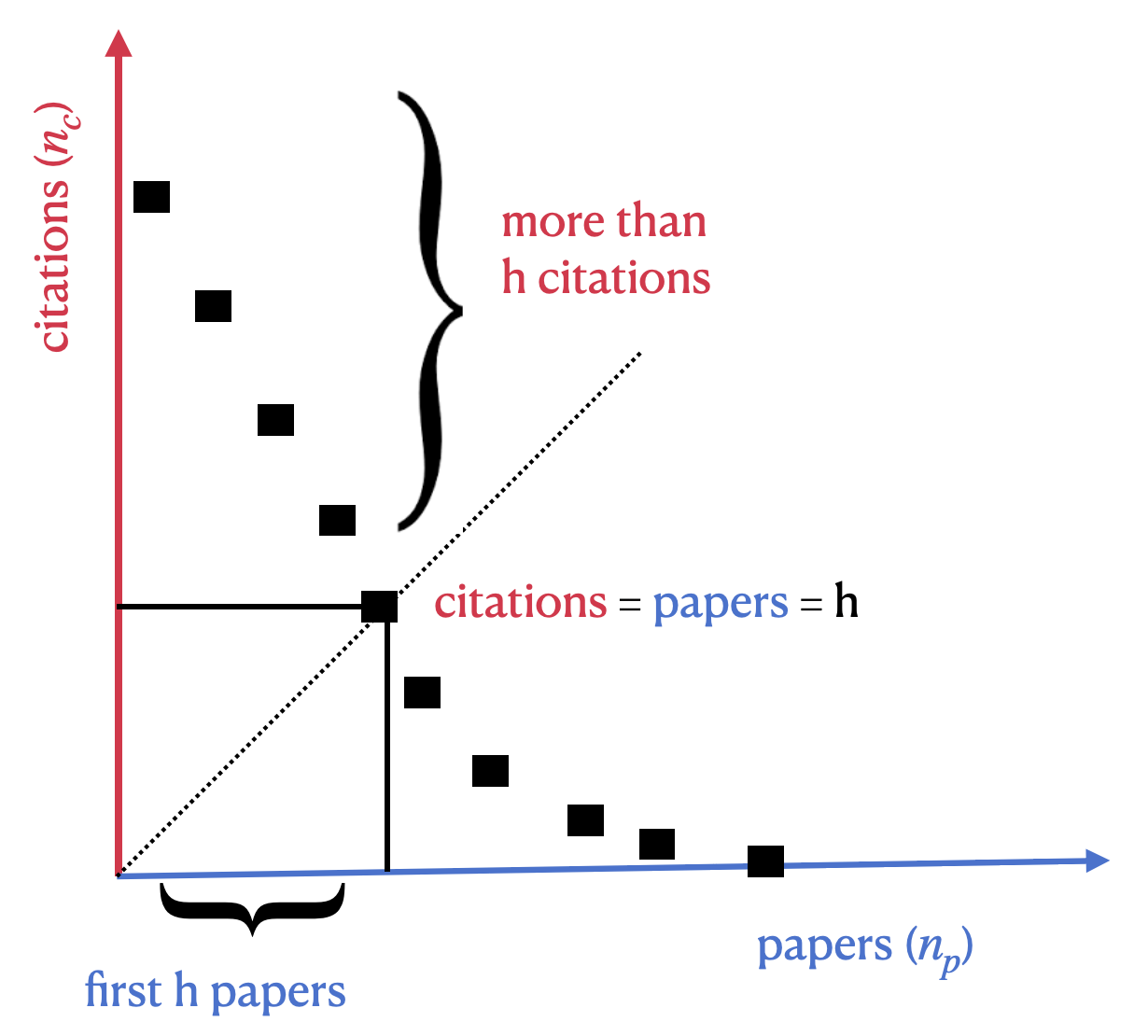}
    \caption{A schematic variation of the number of citations ($n_c$) vs. the number of papers ($n_p$), each having $n_c$ citations, by an author ($\sum_p n_p=N_t$, $\sum_p n_c=C_t$ ). The $45^0$ line from the origin cuts the nonlinear $n_c(n_p)$  line at the ordinate value $h$, giving the number of $h$ papers by the author each of which has earned at least $h$ citations.}
    \label{fig2}
\end{figure}

\section{Inequality measuring indices}
\subsection{Gini and Kolkata Indices}
Quantitative measurements of economic inequalities, specifically the income or wealth inequalities, in different societies or countries started with the introduction of Lorenz function \cite{10Lorenz1905} in 1905. The Lorenz function or curve is represented graphically by plotting the cumulative fraction of the income/wealth or citations   ($L(x)$) earned or possessed against the fraction $x$ of the poorest individuals or papers, when the population or number of papers by any author is arranged in ascending order of their income or citation; see Fig. \ref{fig1}. When everyone in the country earns the same or every paper by an author earns the same number of citations, the Lorenz curve becomes the equality line represented by the dotted-dashed (black) diagonal in Fig. \ref{fig1}. Because of the inequalities in wealth or citations among the agents of a country or the papers of any author, the typical nonlinear shape of the Lorenz curve will have the form of the red line (with  $L(0) = 0$ and $L(1) = 1$). Although $L(x)$ contains all the information of inequality, typical values of some inequality indices help summarizing or characterizing the nature of inequality in the society. One of the oldest and still most popular one is the Gini index \cite{3Gini1912} ($g$) given by the normalized area between the equality line and the Lorenz curve (see Fig. \ref{fig1}). If we define the complementary Lorenz function ($L_c(x) \equiv 1 - L(x)$) , shown in green in Fig. \ref{fig1}, the Kolkata index ($k$) is determined by the fixed point of the complementary Lorenz curve: $L_c(k) = k$ or $L(k) = 1 - k$. Geometrically, it gives the ordinate point at which the Lorenz curve intersects the diagonal line perpendicular to the equality line, giving the fraction $k$ of wealth (citations) that is possessed by the richest $1-k$ fraction of people (papers). Note, $k = 0.80$ corresponds to Pareto’s 80-20 law and, as such, the $k$ index generalizes
the Pareto law.

\subsection{Hirsch Index}
The index for measuring the success inequality among the authors or scientists, the Hirsch index, was proposed \cite{1Hirsch05} in 2005. It is obtained (see Fig. \ref{fig2})  from the corresponding  fixed point of the nonlinear (Ziff-like) citation function $n_c$ giving the number $n$ of papers by an author having citation $c:n_c(h) = h$. This fixed point value $h$ gives the number $h$  of papers by the author, each of which has earned at least $h$ citations.

\begin{figure}[!tbh]
    \centering
    \includegraphics[width=1.0\linewidth]{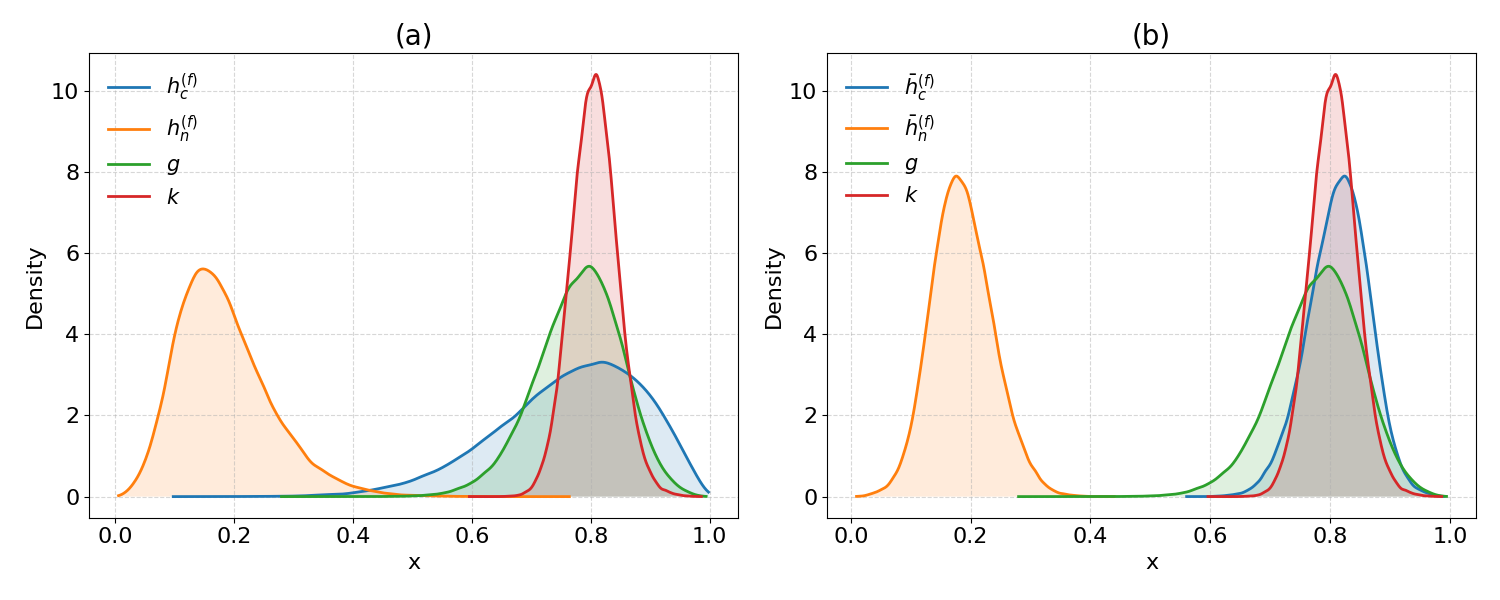}
    \caption{Probability-density distributions of the inequality measures for the Google Scholar sample of 126,067 scientists. (a) Shows the distributions of $h_c^{(f)}$, $h_n^{(n)}$, $g$ and $k$. (b) Shows the corresponding distributions of $\bar{f}_c^{(f)}$, $\bar{h}_n^{(f)}$, $g$ and $k$.}
    \label{fig3}
\end{figure}

\subsection{Normalized Fractional  Hirsch Index}
As we mentioned earlier, while both Gini ($g$) and Kolkata ($k$) indices for the citation inequalities of successful researchers, like the Nobel Laureates, tend to converge towards some universal values ($g = k \simeq 0.86$, higher than the Pareto value $k = 0.80$; see e.g., \cite{5Ghosh14}, \cite{7Banerjee23} and \cite{9Biswas25}), the Hirsch index ($h$) value differs  by orders of magnitude (see e.g., \cite{9Biswas25}). As $h$ index has integer values and although it is generally higher for the successful authors, statistically it can be argued to grow \cite{11Yong14,12Ghosh23} with the square root of  the total number $N_t = \sum_p n_p$ of publications (set denoted by $p$) of the author. The $h$-index value therefore does not have any upper bound.

In an attempt to reconcile, we first define two fractional quantities: $h_n^{(f)} = (h/N_t )$ representing the fraction of successful $h$ papers of the author compared to the total $N_t$ papers published by the author and $h_c^{(f)} = \sum_h n_c/ C_t$  (with $C_t = \sum_p n_c$),  representing the citation fraction of the $h$ successful papers compared to the total number citations $C_t$ from all the papers by the same author. Noting that since statistically $h$ scales with the square root of $C_t$ or $N_t$ (see e.g., \cite{11Yong14,12Ghosh23}),  both $h_n^{(f)}$ and  $h_c^{(f))}$ scales as $1/\sqrt{N_t}$.  Hence we finally define the normalized fraction of citations of the $h$ successful papers $\bar{h}_c^{(f)} =h_c^{(f))}/[h_c^{(f)} + h_n^{(f)}]$  earned by the normalized fraction $\bar{h}_n^{(f)} = h_n^{(f)} /[ h_c^{(f)} + h_n^{(f)}]$  of the $h$ successful papers. Both these normalized fractions are bounded between 0  and 1, and $\bar{h}_c^{(f)} + \bar{h}_n^{(f)} = 1$ by the way we defined them.

\section{Citation data analysis}
\begin{figure}[!tbh]
    \centering
    \includegraphics[width=1.0\linewidth]{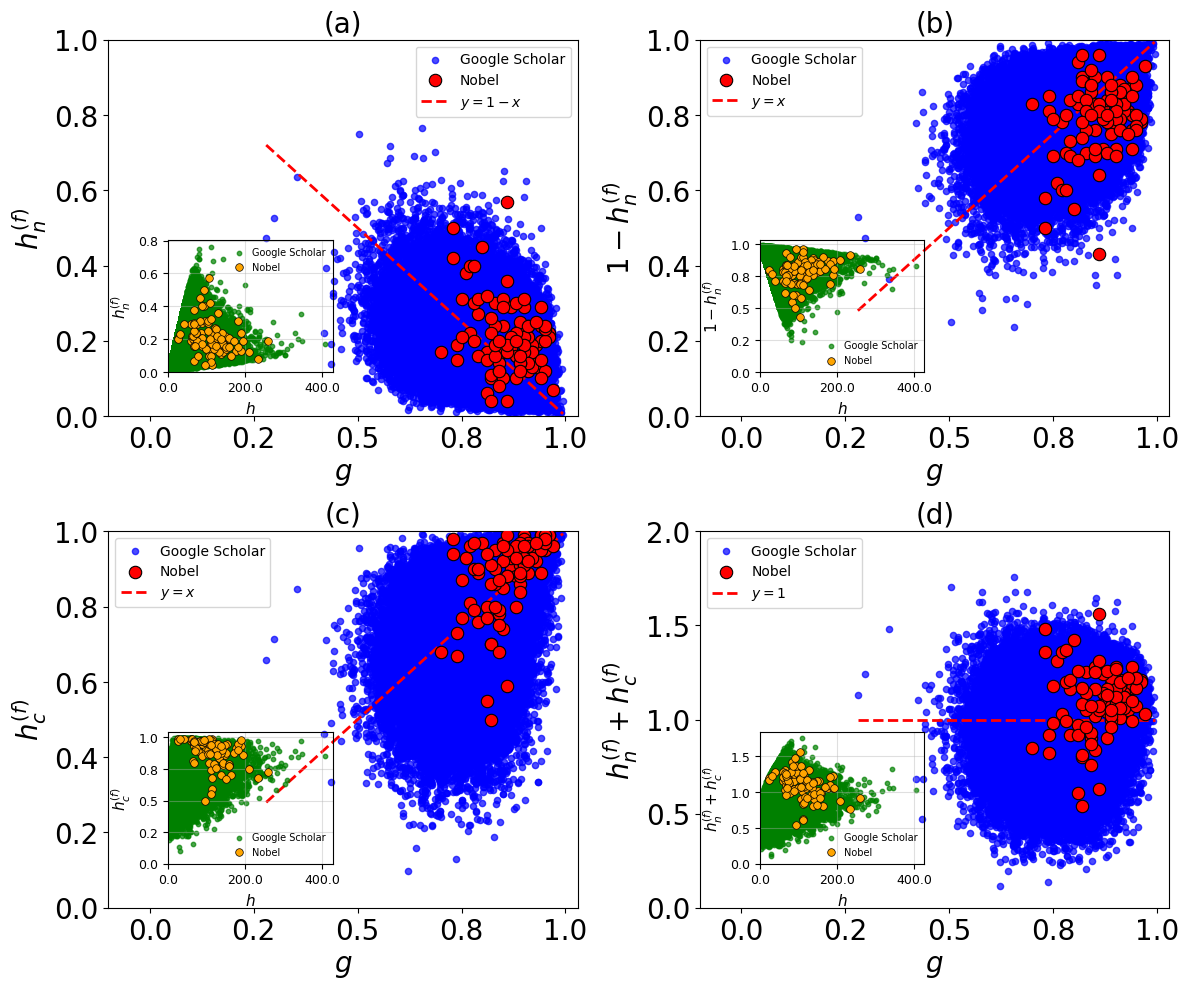}
    \caption{Plots of (a) fraction ($h^{(f)}_n$) of the $h$ successful papers
against the Gini index ($g$) in the main Fig.,  and against Hirsch
index $h$ in the inset, (b)  fraction (1 - $h^{(f)}_n$) for the $h$
successful papers against the Gini index ($g$) in the main Fig., and
against Hirsch index $h$ in the inset, (c) citation  fraction
($h^{(f)}_c$) of the $h$ successful papers against the Gini index
($g$) in the main Fig., and against Hirsch index $h$ in the inset,
(d) ($h^{(f)}_c + h^{(f)}_n$) against Gini index ($g$) in the 
main Fig., and against $h$ in the inset for all the 126,067 scientists
and of the 95 Nobel Laureates (indicated using different colours).}
    \label{fig4}
\end{figure}

\begin{figure}[!tbh]
    \centering
    \includegraphics[width=1.0\linewidth]{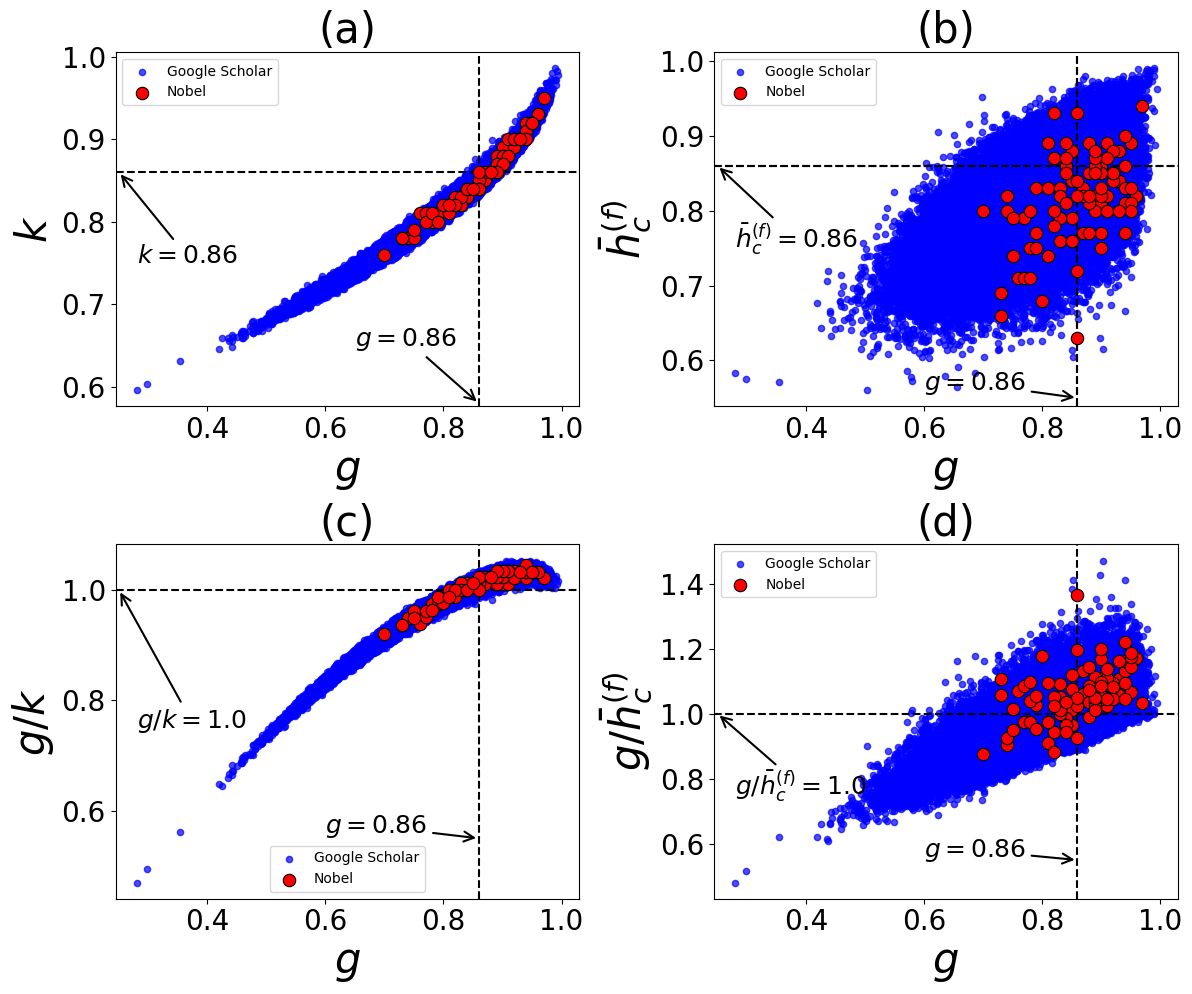}
    \caption{Variations  of  (a) the Kolkata index ($k$)  and (b) the normalized
Hirsch index fraction of citations ($\bar{h}{(f)}_c$) against the Gini
index $g$  for all the 126,067 scientists and of the 95 Nobel
Laureates (indicated using different colours). The location of $k = g
= 0.86$ and of   $\bar{h}^{(f)}_c = g = 0.86$ are indicated. Variations of
(c) $g/k$ against $g$ and of $g/\bar{h}^{(f)}_c$  against  $g$ are shown,
with the locations of $g/k = 1$ and of $g/\bar{h}^{(f)}_c = 1$  at  $g =
0.86$ are shown  for all 126,067 scientists and for the 95 Nobel
Laureates (indicated using different colours).}
    \label{fig5}
\end{figure}

We now compute them,  by analyzing the  Google Scholar data, for 126,067 scientists (including 95 Nobel Laureates) each of whom has at least 100 papers (to have sufficient statistics for evaluating the citation inequality Lorenz curve and the corresponding values of Gini $g$ and Kolkata $k$ indices.  We  evaluate $h_c^{(f)}$, $h_n^{(f)}$, and also $\bar{h}_c^{(f)}$, as well as $\bar{h}_n^{(f)}$ (see Figs. \ref{fig3},\ref{fig4},\ref{fig5}) including those $h$ for the 95 Nobel Laureates (see Figs. \ref{fig3},\ref{fig4},\ref{fig5} and the Tables I-IV). One can see  how their $g$ and $k$ values converge towards a value about 0.86  for the successful scientists, and also their $\bar{h}_c^{(f)}$ values compare with $k$ (and their $\bar{h}_n^{(f)}$ compare with 1 - $k$ values), converging respectively to values around 0.86 and 0.14  for the successful scientists (Nobel Laureates in particular). As may be noted, these observations fit to a modified Pareto's 80-20 law.

\section{Summary and Conclusion}
The Hirsch index ($h$), introduced in 2005 \cite{1Hirsch05}, has since become a very popular inequality metric for the measurement of publication inequalities among scientific authors, and the value of this index for scientific authors is now regularly available in various databases like Google Scholar and Web of Science. Although in general  more successful authors have higher $h$ values, they are not very characteristic of the scientists, and it has been argued \cite{11Yong14} and observed \cite{12Ghosh23} to grow with the total number of publications $N_t$ by the author.  Indeed, as may be seen from Tables I-IV, the citation statistics of 95 Nobel Laureates in Economics, Physics, Chemistry and Medicine (during 2005-2025) have wide variations in the range $25 \le h \le 260$.

Economic inequalities like the Gini index ($g$), introduced in 1912 \cite{3Gini1912} (giving an overall measure of wealth or income inequality of any country) or the Kolkata index ($k$), introduced in 2014 \cite{5Ghosh14} (giving the fraction $k$ of wealth or income possessed by the richest $(1 -k$) fraction of population) are normalized and bounded: $0 \le g \le 1$ and $1/2 \le k \le 1$.  As observed earlier \cite{Ghosh26}  from the analysis of the IRS US income data or some revenue data from movie income, as the competition becomes fierce, and the social welfare supports are either gradually withdrawn or are absent, $g$ and $k$ values approach each other and they become equal to about 0.86 ($k$ = 0.80 corresponds the case of $80\%$  wealth possessed by $20\%$ of the richest; the Pareto's 80-20 law \cite{6Pareto1896}).  It may be mentioned here that in the self-organised critical (SOC) sandpile models, if one plots the fraction of cumulative avalanche mass against the fraction of the number of avalanches (from the smallest to the highest in size) in the sandpile to get the Lorenz curve  (Fig. \ref{fig1})  and determine the Gini $g$ and Kolkata $k$ indices as the sandpile grows towards the respective SOC point,  one gets $g = k \simeq 0.86$, for a number of sandpile models with different dynamics and SOC points \cite{8Manna22}.

In view of these, we propose to define a normalized fractional version $ \bar{h}_c^{(f)}$ of the Hirsch index $h$, and compare its variation and saturation value with that for the Kolkata index $k$ for the citation inequalities of all the 126,067 scientists, including 95 Nobel Laureates  (see Fig. \ref{fig5}). In order to get this quantity  $\bar{h}_c^{(f)}$, we first find the fractional value of the $h$ index given by $h_n^{(f)} = (h/N_t )$ representing the fraction of successful $h$ papers of the author compared to the total $N_t$ ($N_t=\sum_p n_p$) papers published by the author and  also $h_c^{(f)} = \sum_h n_c/ C_t$  (with $C_t = \sum_p n_c $),  representing the citation fraction of the $h$ successful papers compared to the total number of citations $C_t$ from all the papers by the same author. Plots of $h_n^{(f)} $ (in Fig.  \ref{fig4}a) and of  $h_c^{(f)}$ (in Fig. \ref{fig4}c) against $g$ (in the main Figs.) and against $h$ (in the insets) are shown for all scientists considered here. However as $h$ scales with $\sqrt N_t$,  large scatters are still seen as both $h_n^{(f)}$ and $h_c^{(f)}$  remain unnormalized and scale with the inverse of $\sqrt N_t$.  We then evaluated  $\bar{h}_c^{(f)}=  h_c^{(f)} /[ h_c^{(f)} + h_n^{(f)}]$   as the normalized fractional $h$  (comparable to the $k$ index) and its complement  $\bar{h}_n^{(f)} = h_n^{(f)} /[ h_c^{(f)} + h_n^{(f)}]$  (comparable to $1-k$). Variations of $ \bar{h}_c^{(f)}$  and of  $g/\bar{h}_c^{(f)}$  with $g$ are shown in  Figs. \ref{fig5}b and \ref{fig5}d, compared with those for $k$ and $g/k$ against $g$ values (shown in Figs. \ref{fig5} a and \ref{fig5}c) for all the 95 Nobel Laureates against the backdrop of those for all the 126,067 scientists are shown.

Our analysis here suggests that, though the bare (unnormalized) Hirsch index does not show any characteristic value for the distinguished scientists (95 Nobel Laureates considered here have $25 \le h \le 260$, see tables I-IV), the normalized fractional Hirsch index $ \bar{h}_c^{(f)}$ have much more restricted range ($0.68 \le \bar{h}_c^{(f)} \le 0.98$ for all the 95 Nobel Laureates; see Tables I-IV).  This is similar to the observation (see Figs. \ref{fig5}c and \ref{fig5}d) of $g = k \simeq 0.86$ and $g = \bar{h}_c^{(f)} \simeq 0.86$ for 95 Nobel Laureates. As mentioned earlier,  this value of the Kolkata index $k$ or of the normalized fractional Hirsch index $ \bar{h}_c^{(f)}$ (though a little above the inequality value corresponding to Pareto's 80-20 law) seems to correspond well to the competitive Self-Organised dynamical system's critical value \cite{8Manna22,9Biswas25}  of inequality level ($k\simeq  \bar{h}_c^{(f)} \simeq 0.86$).

\begin{acknowledgments}
We are grateful to the AI Journal ``gist.science" for publishing \cite{gist} a
widely readable summary of the study reported in the paper
(\url{arXiv:2609.26710v1}).
\end{acknowledgments}

\appendix
\begin{table*}[htbp]
\centering
\caption{Hirsch index ($h$),  total papers ($N_t$), total citations ($C_h$)
received by the top cited $h$ paper, total citations ($C_t$) received
by all the $N_t$ papers , fractional citations ($h^{(f)}_c$)  of the
$h$ successful papers,  fraction ($h^{(f)}_n$) of the $h$ successful
papers, normalized Hirsch index fraction of citations ($\bar{h}{(f)}_c$),
normalized Hirsch index of papers ($\bar{h}^{(f)}_n$), Gini index ($g$) and
Kolkata index ($k$) for the papers published by each of the Nobel
laureates in economic science (collected for years 2005–2025, each
having more than 100 publications; from the Google Scholar database).}
\begin{tabular}{|l|l|r|r|r|r|r|r|r|r|r|r|}
\hline
Year & Name & $h$ &  $N_t$ & $C_h$ & $C_t$ & $h_{c}^{(f)}$ &$h_{n}^{(f)}$ & $\bar{h}_{c}^{(f)}$ & $\bar{h}_{n}^{(f)}=$ & $g$ & $k$ \\
& &  &  &  &  & $=C_h/C_t$ & $=h/N_t$ & & $1-\bar{h}_{c}^{(f)}$ & &  \\
\hline
\hline
2005 & Robert J. Aumann & 69 & 306 & 42766 & 44615 & 0.96 & 0.23 & 0.81 & 0.19 & 0.89 & 0.88 \\
2007 & Roger B. Myerson & 60 & 317 & 49592 & 51277 & 0.97 & 0.19 & 0.84 & 0.16 & 0.93 & 0.90 \\
2009 & Elinor Ostrom & 176 & 1343 & 275425 & 297924 & 0.92 & 0.13 & 0.88 & 0.12 & 0.93 & 0.90 \\
2010 & Christopher A. Pissarides & 68 & 425 & 40284 & 42840 & 0.94 & 0.16 & 0.85 & 0.15 & 0.92 & 0.89 \\
2010 & Peter A. Diamond & 66 & 172 & 27782 & 30003 & 0.93 & 0.38 & 0.71 & 0.29 & 0.76 & 0.81 \\
2011 & Christopher A. Sims & 84 & 347 & 82459 & 85340 & 0.97 & 0.24 & 0.80 & 0.20 & 0.91 & 0.88 \\
2011 & Thomas J. Sargent & 109 & 941 & 69649 & 77982 & 0.89 & 0.12 & 0.89 & 0.11 & 0.91 & 0.89 \\
2012 & Alvin E. Roth & 112 & 615 & 62316 & 69260 & 0.90 & 0.18 & 0.83 & 0.17 & 0.87 & 0.85 \\
2012 & Lloyd S. Shapley & 65 & 328 & 69582 & 71330 & 0.97 & 0.20 & 0.83 & 0.17 & 0.94 & 0.91 \\
2013 & Eugene F. Fama & 114 & 476 & 419457 & 424209 & 0.99 & 0.24 & 0.81 & 0.19 & 0.94 & 0.91 \\
2013 & Lars Peter Hansen & 78 & 419 & 61637 & 65123 & 0.95 & 0.19 & 0.84 & 0.16 & 0.92 & 0.89 \\
2014 & Jean Tirole & 153 & 731 & 203089 & 213971 & 0.95 & 0.21 & 0.82 & 0.18 & 0.89 & 0.87 \\
2015 & Angus Deaton & 124 & 612 & 124432 & 131215 & 0.95 & 0.20 & 0.82 & 0.18 & 0.90 & 0.87 \\
2016 & Bengt Holmstrom & 60 & 204 & 87864 & 88805 & 0.99 & 0.29 & 0.77 & 0.23 & 0.90 & 0.88 \\
2017 & Richard H. Thaler & 111 & 474 & 232083 & 237769 & 0.98 & 0.23 & 0.81 & 0.19 & 0.90 & 0.89 \\
2018 & Paul M. Romer & 61 & 280 & 140518 & 142326 & 0.99 & 0.22 & 0.82 & 0.18 & 0.96 & 0.93 \\
2018 & William D. Nordhaus & 132 & 820 & 110819 & 122133 & 0.91 & 0.16 & 0.85 & 0.15 & 0.89 & 0.87 \\
2019 & Abhijit V. Banerjee & 121 & 815 & 101864 & 111467 & 0.91 & 0.15 & 0.86 & 0.14 & 0.90 & 0.88 \\
2019 & Esther Duflo & 112 & 671 & 119487 & 125319 & 0.95 & 0.17 & 0.85 & 0.15 & 0.92 & 0.89 \\
2019 & Michael Kremer & 163 & 1573 & 83669 & 118732 & 0.70 & 0.10 & 0.87 & 0.13 & 0.82 & 0.82 \\
2020 & Paul R. Milgrom & 85 & 399 & 118211 & 121570 & 0.97 & 0.21 & 0.82 & 0.18 & 0.91 & 0.90 \\
2021 & David Card & 118 & 720 & 100222 & 107679 & 0.93 & 0.16 & 0.85 & 0.15 & 0.90 & 0.88 \\
2021 & Guido W. Imbens & 104 & 418 & 123363 & 127617 & 0.97 & 0.25 & 0.80 & 0.20 & 0.89 & 0.88 \\
2022 & Ben Bernanke & 120 & 717 & 145025 & 152333 & 0.95 & 0.17 & 0.85 & 0.15 & 0.92 & 0.90 \\
2022 & Philip H. Dybvig & 40 & 139 & 57369 & 57994 & 0.99 & 0.29 & 0.77 & 0.23 & 0.94 & 0.92 \\
2023 & Claudia Goldin & 88 & 424 & 56203 & 59089 & 0.95 & 0.21 & 0.82 & 0.18 & 0.90 & 0.87 \\
2024 & Daron Acemoglu & 190 & 1450 & 280361 & 301658 & 0.93 & 0.13 & 0.88 & 0.12 & 0.92 & 0.90 \\
2024 & James A. Robinson & 105 & 876 & 133310 & 140884 & 0.95 & 0.12 & 0.89 & 0.11 & 0.95 & 0.92 \\
2024 & Simon Johnson & 68 & 1023 & 95986 & 100058 & 0.96 & 0.07 & 0.94 & 0.06 & 0.97 & 0.95 \\
2025 & Joel Mokyr & 73 & 440 & 30582 & 33647 & 0.91 & 0.17 & 0.85 & 0.15 & 0.90 & 0.87 \\
2025 & Philippe Aghion & 147 & 676 & 145035 & 155100 & 0.94 & 0.22 & 0.81 & 0.19 & 0.88 & 0.86 \\
\hline
& \multicolumn{5}{c|}{\textbf{Average $\Longrightarrow$}}& 0.94 & 0.19 & 0.83 & 0.17 & 0.91 & 0.89  \\
\hline
\end{tabular}
\label{tab:econ_nobel_metrics}
\end{table*}

\begin{table*}[htbp]
\centering
\caption{Hirsch index ($h$),  total papers ($N_t$), total citations ($C_h$)
received by the top cited $h$ paper, total citations ($C_t$) received
by all the $N_t$ papers , fractional citations ($h^{(f)}_c$)  of the
$h$ successful papers,  fraction ($h^{(f)}_n$) of the $h$ successful
papers, normalized Hirsch index fraction of citations ($\bar{h}{(f)}_c$),
normalized Hirsch index of papers ($\bar{h}^{(f)}_n$), Gini index ($g$) and
Kolkata index ($k$) for the papers published by each of the Nobel
laureates in physics (collected for years 2001–2025, each
having more than 100 publications; from the Google Scholar database)}
\begin{tabular}{|l|l|r|r|r|r|r|r|r|r|r|r|}
\hline
Year & Name & $h$ &  $N_t$ & $C_h$ & $C_t$ & $h_{c}^{(f)}$ &$h_{n}^{(f)}$ & $\bar{h}_{c}^{(f)}$ & $\bar{h}_{n}^{(f)}=$ & $g$ & $k$ \\
& &  &  &  &  & $=C_h/C_t$ & $=h/N_t$ & & $1-\bar{h}_{c}^{(f)}$ & &  \\
\hline
2001 & Wolfgang Ketterle & 125 & 605 & 74711 & 80093 & 0.93 & 0.21 & 0.82 & 0.18 & 0.87 & 0.85 \\
2003 & Anthony J. Leggett & 75 & 400 & 46459 & 50152 & 0.93 & 0.19 & 0.83 & 0.17 & 0.90 & 0.88 \\
2005 & John L. Hall & 88 & 560 & 34131 & 38610 & 0.88 & 0.16 & 0.85 & 0.15 & 0.88 & 0.86 \\
2008 & Makoto Kobayashi & 65 & 301 & 27643 & 33933 & 0.81 & 0.22 & 0.79 & 0.21 & 0.77 & 0.81 \\
2008 & Toshihide Maskawa & 25 & 117 & 20618 & 20795 & 0.99 & 0.21 & 0.82 & 0.18 & 0.96 & 0.93 \\
2010 & Konstantin Novoselov & 181 & 1172 & 430072 & 450549 & 0.95 & 0.15 & 0.86 & 0.14 & 0.94 & 0.91 \\
2012 & Serge Haroche & 99 & 696 & 46829 & 50895 & 0.92 & 0.14 & 0.87 & 0.13 & 0.91 & 0.88 \\
2014 & Hiroshi Amano & 113 & 2618 & 39417 & 66948 & 0.59 & 0.04 & 0.93 & 0.07 & 0.86 & 0.84 \\
2016 & David J. Thouless & 71 & 297 & 61974 & 64696 & 0.96 & 0.24 & 0.80 & 0.20 & 0.91 & 0.88 \\
2016 & F. Duncan M. Haldane & 74 & 255 & 52445 & 55041 & 0.95 & 0.29 & 0.77 & 0.23 & 0.87 & 0.86 \\
2017 & Barry C. Barish & 148 & 1057 & 139173 & 161026 & 0.86 & 0.14 & 0.86 & 0.14 & 0.89 & 0.86\\
2018 & Arthur Ashkin & 66 & 223 & 53292 & 55245 & 0.96 & 0.30 & 0.77 & 0.23 & 0.88 & 0.86 \\
2018 & Donna Strickland & 25 & 125 & 13954 & 14346 & 0.97 & 0.20 & 0.83 & 0.17 & 0.95 & 0.92 \\
2018 & Gerard Mourou & 106 & 1042 & 50129 & 62396 & 0.80 & 0.10 & 0.89 & 0.11 & 0.88 & 0.86 \\
2019 & Didier Queloz & 138 & 1264 & 54060 & 76900 & 0.70 & 0.11 & 0.87 & 0.13 & 0.82 & 0.82 \\
2021 & Giorgio Parisi & 141 & 1146 & 89831 & 114613 & 0.78 & 0.12 & 0.86 & 0.14 & 0.84 & 0.83\\
2021 & Syukuro Manabe & 88 & 295 & 47793 & 50075 & 0.95 & 0.30 & 0.76 & 0.24 & 0.83 & 0.83 \\
2022 & Alain Aspect & 77 & 775 & 38899 & 43496 & 0.89 & 0.10 & 0.90 & 0.10 & 0.94 & 0.90 \\
2022 & Anton Zeilinger & 150 & 1098 & 111257 & 125838 & 0.88 & 0.14 & 0.87 & 0.13 & 0.89 & 0.87 \\
2022 & John F. Clauser & 31 & 135 & 23038 & 23355 & 0.99 & 0.23 & 0.81 & 0.19 & 0.95 & 0.92 \\
2023 & Anne L'Huillier & 88 & 531 & 32009 & 37858 & 0.85 & 0.17 & 0.84 & 0.16 & 0.85 & 0.84 \\
2023 & Ferenc Krausz & 132 & 1124 & 77982 & 93238 & 0.84 & 0.12 & 0.88 & 0.12 & 0.89 & 0.86 \\
2024 & Geoffrey E. Hinton & 190 & 787 & 1026400 & 1047259 & 0.98 & 0.24 & 0.80 & 0.20 & 0.95 & 0.92 \\
2024 & John J. Hopfield & 96 & 306 & 93592 & 97098 & 0.96 & 0.31 & 0.75 & 0.25 & 0.90 & 0.87 \\
2025 & John Clarke & 111 & 1105 & 38405 & 52087 & 0.74 & 0.10 & 0.88 & 0.12 & 0.85 & 0.84 \\
2025 & Michel H. Devoret & 136 & 781 & 74765 & 84599 & 0.88 & 0.17 & 0.84 & 0.16 & 0.86 & 0.85 \\
\hline
& \multicolumn{5}{c|}{\textbf{Average $\Longrightarrow$}}& 0.88 & 0.18 & 0.84 & 0.16 & 0.87 & 0.87  \\
\hline
\end{tabular}
\label{tab:econ_nobel_metrics}
\end{table*}

\begin{table*}[htbp]
\centering
\caption{Hirsch index ($h$),  total papers ($N_t$), total citations ($C_h$)
received by the top cited $h$ paper, total citations ($C_t$) received
by all the $N_t$ papers , fractional citations ($h^{(f)}_c$)  of the
$h$ successful papers,  fraction ($h^{(f)}_n$) of the $h$ successful
papers, normalized Hirsch index fraction of citations ($\bar{h}{(f)}_c$),
normalized Hirsch index of papers ($\bar{h}^{(f)}_n$), Gini index ($g$) and
Kolkata index ($k$) for the papers published by each of the Nobel
laureates in Chemistry (collected for years 2001–2025, each
having more than 100 publications; from the Google Scholar database)}
\begin{tabular}{|l|l|r|r|r|r|r|r|r|r|r|r|}
\hline
Year & Name & $h$ &  $N_t$ & $C_h$ & $C_t$ & $h_{c}^{(f)}$ &$h_{n}^{(f)}$ & $\bar{h}_{c}^{(f)}$ & $\bar{h}_{n}^{(f)}=$ & $g$ & $k$ \\
& &  &  &  &  & $=C_h/C_t$ & $=h/N_t$ & & $1-\bar{h}_{c}^{(f)}$ & &  \\
\hline
\hline
2001 & Ryoji Noyori & 141 & 840 & 73937 & 92189 & 0.80 & 0.17 & 0.83 & 0.17 & 0.81 & 0.82 \\
2002 & Koichi Tanaka & 111 & 1735 & 25508 & 46786 & 0.55 & 0.06 & 0.89 & 0.11 & 0.81 & 0.82 \\
2003 & Peter Agre & 121 & 456 & 52269 & 58715 & 0.89 & 0.27 & 0.77 & 0.23 & 0.79 & 0.81 \\
2007 & Gerhard Ertl & 149 & 1004 & 59597 & 88362 & 0.67 & 0.15 & 0.82 & 0.18 & 0.74 & 0.78 \\
2010 & Akira Suzuki & 95 & 2386 & 26188 & 52653 & 0.50 & 0.04 & 0.93 & 0.07 & 0.82 & 0.82 \\
2012 & Robert J. Lefkowitz & 260 & 1391 & 162565 & 221859 & 0.73 & 0.19 & 0.80 & 0.20 & 0.74 & 0.78 \\
2013 & Michael Levitt & 103 & 342 & 45065 & 50241 & 0.90 & 0.30 & 0.75 & 0.25 & 0.78 & 0.80 \\
2014 & Eric Betzig & 82 & 266 & 50446 & 52183 & 0.97 & 0.31 & 0.76 & 0.24 & 0.85 & 0.84 \\
2014 & Stefan W. Hell & 158 & 786 & 75209 & 94702 & 0.79 & 0.20 & 0.80 & 0.20 & 0.78 & 0.80 \\
2014 & William E. Moerner & 118 & 827 & 45244 & 57374 & 0.79 & 0.14 & 0.85 & 0.15 & 0.84 & 0.83 \\
2016 & Jean-Pierre Sauvage & 114 & 661 & 39139 & 57312 & 0.68 & 0.17 & 0.80 & 0.20 & 0.70 & 0.76 \\
2017 & Joachim Frank & 123 & 769 & 43559 & 57033 & 0.76 & 0.16 & 0.83 & 0.17 & 0.79 & 0.81 \\
2017 & Richard Henderson & 72 & 294 & 33068 & 35862 & 0.92 & 0.24 & 0.79 & 0.21 & 0.85 & 0.84 \\
2018 & Frances H. Arnold & 157 & 750 & 66709 & 86413 & 0.77 & 0.21 & 0.79 & 0.21 & 0.75 & 0.78 \\
2019 & John B. Goodenough & 209 & 1796 & 169819 & 225462 & 0.75 & 0.12 & 0.87 & 0.13 & 0.84 & 0.83 \\
2020 & Emmanuelle Charpentier & 67 & 269 & 67441 & 69143 & 0.97 & 0.25 & 0.80 & 0.20 & 0.93 & 0.90 \\
2020 & Jennifer A. Doudna & 170 & 894 & 146777 & 167456 & 0.88 & 0.19 & 0.82 & 0.18 & 0.86 & 0.84 \\
2021 & Benjamin List & 104 & 337 & 43499 & 50065 & 0.87 & 0.31 & 0.74 & 0.26 & 0.75 & 0.79 \\
2021 & David W. C. MacMillan & 139 & 581 & 82607 & 91499 & 0.90 & 0.24 & 0.79 & 0.21 & 0.83 & 0.83 \\
2022 & Carolyn R. Bertozzi & 157 & 1034 & 84036 & 109762 & 0.77 & 0.15 & 0.83 & 0.17 & 0.81 & 0.81 \\
2022 & Morten Meldal & 68 & 425 & 26324 & 32976 & 0.80 & 0.16 & 0.83 & 0.17 & 0.83 & 0.82 \\
2023 & Louis E. Brus & 123 & 471 & 86785 & 95104 & 0.91 & 0.26 & 0.78 & 0.22 & 0.82 & 0.83 \\
2023 & Moungi G. Bawendi & 193 & 1020 & 158056 & 184242 & 0.86 & 0.19 & 0.82 & 0.18 & 0.83 & 0.83 \\
2024 & David Baker & 233 & 2788 & 148408 & 216930 & 0.68 & 0.08 & 0.89 & 0.11 & 0.84 & 0.84 \\
2024 & Demis Hassabis & 105 & 184 & 284296 & 286680 & 0.99 & 0.57 & 0.63 & 0.37 & 0.86 & 0.86 \\
\hline
& \multicolumn{5}{c|}{\textbf{Average $\Longrightarrow$}}& 0.80 & 0.21 & 0.81 & 0.19 & 0.81 & 0.82  \\
\hline
\end{tabular}
\label{tab:econ_nobel_metrics}
\end{table*}

\begin{table*}[htbp]
\centering
\caption{ Hirsch index ($h$),  total papers ($N_t$), total citations ($C_h$)
received by the top cited $h$ paper, total citations ($C_t$) received
by all the $N_t$ papers , fractional citations ($h^{(f)}_c$)  of the
$h$ successful papers,  fraction ($h^{(f)}_n$) of the $h$ successful
papers, normalized Hirsch index fraction of citations ($\bar{h}{(f)}_c$),
normalized Hirsch index of papers ($\bar{h}^{(f)}_n$), Gini index ($g$) and
Kolkata index ($k$) for the papers published by each of the Nobel
laureates in Physiology or Medicine (collected for years 2001–2025, each
having more than 100 publications; from the Google Scholar database)}
\begin{tabular}{|l|l|r|r|r|r|r|r|r|r|r|r|}
\hline
Year & Name & $h$ &  $N_t$ & $C_h$ & $C_t$ & $h_{c}^{(f)}$ &$h_{n}^{(f)}$ & $\bar{h}_{c}^{(f)}$ & $\bar{h}_{n}^{(f)}=$ & $g$ & $k$ \\
& &  &  &  &  & $=C_h/C_t$ & $=h/N_t$ & & $1-\bar{h}_{c}^{(f)}$ & &  \\
\hline
\hline
2005 & Barry J. Marshall & 70 & 448 & 32159 & 36340 & 0.89 & 0.16 & 0.85 & 0.15 & 0.89 & 0.86 \\
2009 & Carol W. Greider & 83 & 185 & 60579 & 62368 & 0.97 & 0.45 & 0.68 & 0.32 & 0.80 & 0.82 \\
2009 & Jack W. Szostak & 127 & 574 & 67359 & 78152 & 0.86 & 0.22 & 0.80 & 0.20 & 0.82 & 0.82 \\
2012 & Shinya Yamanaka & 128 & 351 & 138206 & 145402 & 0.95 & 0.36 & 0.72 & 0.28 & 0.86 & 0.84 \\
2014 & Edvard I. Moser & 94 & 238 & 64694 & 67663 & 0.96 & 0.40 & 0.71 & 0.29 & 0.77 & 0.80 \\
2014 & May-Britt Moser & 87 & 216 & 55176 & 57014 & 0.97 & 0.40 & 0.71 & 0.29 & 0.78 & 0.81 \\
2019 & William G. Kaelin & 108 & 256 & 56567 & 59838 & 0.94 & 0.42 & 0.69 & 0.31 & 0.73 & 0.78 \\
2020 & Michael Houghton & 108 & 546 & 54582 & 62746 & 0.87 & 0.20 & 0.81 & 0.19 & 0.84 & 0.84 \\
2021 & Ardem Patapoutian & 92 & 185 & 55774 & 57013 & 0.98 & 0.50 & 0.66 & 0.34 & 0.73 & 0.78 \\
2022 & Svante Pääbo & 182 & 595 & 143836 & 159837 & 0.90 & 0.31 & 0.75 & 0.25 & 0.79 & 0.80 \\
2023 & Katalin Karikó & 67 & 228 & 33999 & 35536 & 0.96 & 0.29 & 0.76 & 0.24 & 0.85 & 0.84 \\
2024 & Gary Ruvkun & 112 & 347 & 70358 & 75262 & 0.94 & 0.32 & 0.74 & 0.26 & 0.81 & 0.82 \\
2025 & Shimon Sakaguchi & 137 & 692 & 137630 & 148819 & 0.93 & 0.20 & 0.82 & 0.18 & 0.88 & 0.86 \\
\hline
& \multicolumn{5}{c|}{\textbf{Average $\Longrightarrow$}}& 0.93 & 0.33 & 0.75 & 0.25 & 0.81 & 0.82 \\
\hline
\end{tabular}
\label{tab:econ_nobel_metrics}
\end{table*}
\end{document}